\documentstyle[11pt,newpasp,twoside,psfig]{article}
\begin{document}

\title{Galaxy Formation and Baryonic Dark Matter}
\author{Fran\c{c}oise Combes}
\affil{Observatoire de Paris, 61 Av. de l'Observatoire,
F-75 014, Paris, France}

\setcounter{page}{411}
\index{Combes, F.}

\begin{abstract}
The $\Lambda$CDM scenario to form galaxies encounters many problems
when confronted with observations, namely the prediction of
dark matter cusps in all galaxies, and in particular in
dwarf irregulars, dominated by dark matter, or the low angular momentum
and consequent small size of galaxy disks, or the high predicted number of
small systems. We will consider the hypothesis that the baryonic dark matter
could be in the form of gas and could be continuously involved in the
galaxy formation scenario. Through cold gas accretion all along a galaxy life,
angular momentum could be increased, galaxy disks could be more dominated by
baryons than previously thought, and small systems could merge more
frequently.
\end{abstract}

\section{Introduction: main problems of the $\Lambda$CDM paradigm}

One of the main problems encountered by the CDM-model has already been
debated actively in this conference: the observed dark halo profiles
of DM-dominated galaxies (dwarfs and LSB) are much better fitted
with constant-density cores than with the steep power-law slopes predicted
by CDM simulations (e.g. Navarro, Frenk \& White 1997).
Rotation curves have now sufficiently large spatial resolution
(their dynamical range is even larger than in simulations), and it appears
that it is no longer possible to explain the discrepancy
by observational artefacts (de Blok et al. 2003). While the observed
slope is in average $\sim$ -0.2, recent simulations show
that the predicted slopes are not flatter than -1.5 (Fukushige \& Makino 2003).
The amount of discrepancy is however disputed (e.g.
Swaters et al 2003).

A second problem is the low angular momentum of condensed baryons in CDM models, 
and consequent small predicted radius of galaxy disks, much too small compared
to observed ones (Navarro \& Steinmetz 2000). CDM simulations reproduce
the Tully-Fisher slope and scatter, but not the zero point.
They end up with dark halos too concentrated, and therefore
a dark-to-visible mass ratio much too large
within the observable disk.

A third problem is the high predicted number of small haloes,
a power spectrum for structures advantaging the small-scales,
leading to a large number of substructures that are not observed
(Moore et al. 1999).

Can the hypothesis that a large fraction of
dark baryons are in the form of cold
condensed gas help to solve the problems?
If we believe the primordial nucleosynthesis constraints,
90\% of baryons are dark, and several 
hypothesis have been explored for the nature of these dark baryons;
baryons in compact objects (brown dwarfs, white dwarfs,
black holes) are either not favored by micro-lensing experiments
or suffer major problems
(Lasserre et al. 2000, Alcock et al. 2001, Afonso et al. 2003).
The best hypothesis today is gas, 
either hot gas in the intergalactic and inter-cluster medium,
or cold gas in the vicinity of galaxies (Pfenniger
et al. 1994, Pfenniger \& Combes 1994).
The presence of large amounts of warm and diffuse gas in space
cannot solve the problems, as shown by most $\Lambda$CDM cosmological
simulations based on this hypothesis, so we will
concentrate on the second option.

\section{Cusps in Galaxy centers}

This central problem has been the subject of many efforts
to solve it, either redistributing the dark matter
or changing its nature, but up to now no one is perfectly satisfying:
-- black hole binaries can heat and
flatten cusps (Milosavljevic \& Merritt 2001), but they do
not exist in dwarf galaxies, 
-- stellar feedback (Gnedin \& Zhao 2002),
but even generous amounts of
feedback are insufficient to destroy the central halo cusp, 
while the inner density is lowered only by a modest factor of 2 to 6;
-- bars (Weinberg \& Katz 2002) have only weak
effects and are not likely in dwarf Irr again.
Among the various alternatives to the $\Lambda$CDM model,
warm dark matter WDM appears not promising (Knebe et al 2002),
nor self-interacting dark matter SIDM (Spergel \& Steinhardt 2000).
Changing the cosmological parameters, matter
density or power spectrum, allowing little power
on galaxy scales (McGaugh et al 2003), bring only
partial solutions.

Dwarf Irr galaxies are dominated by dark matter, but also their gas 
mass is dominating their stellar mass. It is remarkable that they 
obey the relation $\sigma_{DM}$/$\sigma_{HI}$ = cste, revealing
that the dark-matter and HI gas surface densities have the
same radial distribution (e.g. Hoekstra et al. 2001).
All rotation curves can be explained, when the observed surface
density of gas is multiplied by a constant factor, around 7-10.
This $\sigma_{DM}$/$\sigma_{HI}$ ratio is almost constant
from galaxy to galaxy, being slightly larger for early-types
(Combes 2002). A solution to the problem of non-observed CDM cusps
is that CDM would not be dominating in LSB galaxy centers, 
as is already the case in more evolved early-type galaxies, 
dominated by the stars. On the contrary, the mass in the center
of LSB HI-dominated galaxies could be predominantly baryonic,
in the form of cold condensed molecular gas. The column density
of this H$_2$ gas would be about one order of magnitude larger
than the HI column density, which is already the case in 
early-type galaxies (where the larger metallicity of the gas
allows to observe the H$_2$ tracer, i.e. the CO molecule).
That the gas mass has an essential dynamical role is supported
by the success of the baryonic Tully-Fisher relation (McGaugh et al. 2000).
It is interesting to note that dark haloes at cluster scale 
are also observed with a core (and not a cusp, Sand et al. 2002), and this
could be due to the mass domination of baryons in the center
(El-Zant et al 2003).

The influence of the baryon fraction $\Omega_b$ on galaxy formation
has been investigated through numerical simulations by Gardner et al. (2003): 
they show that increasing $\Omega_b$ does not change the amount of
diffuse gas in the intergalactic medium, but on the contrary the
condensed phase associated to galaxies increases faster then $\Omega_b$,
due to the increased cooling rate. This leads to a considerably larger
baryon representation in virialized systems with respect to 
their universal fraction. Another factor is the limited spatial
resolution of the simulations, which could reduce the concentration of
the dissipative material.

\section{Angular momentum and disk formation}

Most CDM cosmological simulations encounter a fundamental problem
to form galaxy disks: due to low specific angular momentum for
baryons, disks end up with too small radii (e.g. Abadi et al. 2003).
Similarly elliptical systems end up too concentrated (Meza et al. 2003).
In the commonly adopted paradigm, baryons at the start have the
same specific angular momentum than dark matter. But
then, in the course of evolution, 
baryons lose their angular momentum to the CDM, through
dynamical friction. This is due to the main
formation mechanism of galaxies, through hierarchical 
merging. The problem is the too early concentration of
baryons, and the subsequent
formation of galaxy disks as the final outcome of a sequence
of merger events. Instead disks could be
formed essentially through gas accretion from large-scale filaments,
and this accretion could be prolonged until late times. 

\subsection{Feedback and cold gas accretion}

The main solution to the angular momentum problem has been
to increase the efficiency of feedback processes due to star formation.
Supernovae should provide enough energy in the interstellar
medium to prevent further collapse and star formation, 
and maintain disks with angular momentum close
to the value required to fit observations (Thacker \& Couchman 2001).
In the same vein, Binney et al (2001) propose that, after the formation
of protogalactic disks, massive winds expell considerable amounts of baryons.
This contributes to the expansion of the inner dark matter (and
the formation of cores), while the presently observed disks form late, 
when the dark halo contribution in the inner parts has been reduced.

Weil et al (1998) varied the gas cooling epoch, and found that
discs can form by the present day with correct angular momenta
if radiative cooling is suppressed until z=1.  This means
that feedback processes must be efficient enough to
prevent gas from collapsing until late epochs.
Eke et al. (2000)  delay the gas cooling until z=1, and succeed to form
disks with angular momentum approximately conserved during collapse.
They show however that keeping enough specific angular momentum
depends on the cosmology adopted, on the inhomogeneity of the haloes
after the gas begins to collapse, and  of course on the value of
$\Omega_m$, larger values of $\Lambda$ for flat universes being
much more favourable.

How does the gas collapse? In semi-analytical models (e.g. Mo et al. 1998),
the gas is assumed to be hot and shock heated to the Virial temperature of the halo.
But another way to accrete mass is cold gas mass accretion
(Katz et al. 2002, Binney 2003).
Gas is channeled through filaments, and only moderately heated by weak 
shocks, and radiating quickly. It never heats to the Virial temperature 
of its halo. Accretion is not spherical, gas keeps angular momentum
and rotates near the galaxies, making it more easy to form disks.
Cold mode of gas accretion is the most
efficient at z larger than 1. 

\begin{figure}[t]
\centerline{
\psfig{figure=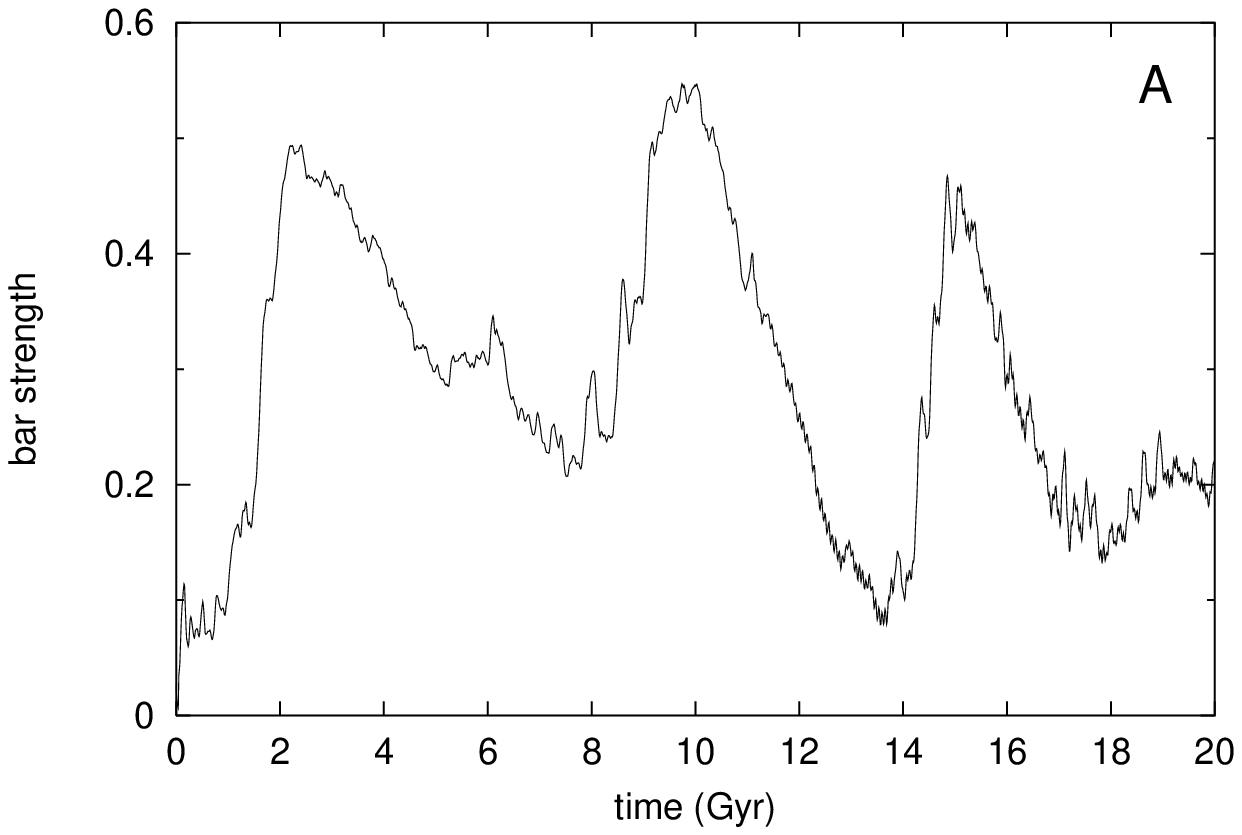,width=6.5cm,angle=0}
\psfig{figure=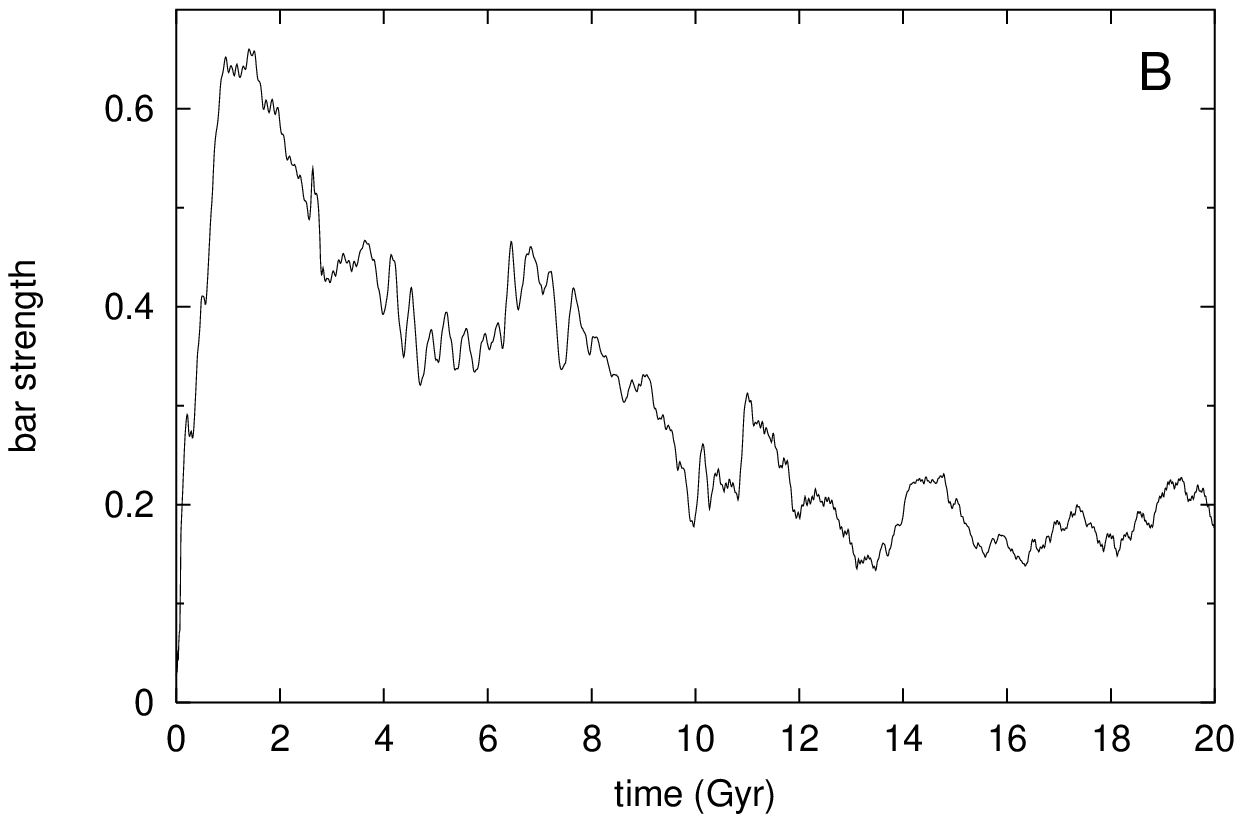,width=6.5cm,angle=0}
}
\caption{Evolution of the bar strength over a Hubble time,
showing that external gas accretion can revive bars in galaxy disks
(numerical simulations from Bournaud \& Combes 2002).}
\label{fig1}
\end{figure}

\subsection{Gas accretion and bar strength}

The bar frequency in nowadays galaxy disks can provide
a way of quantifying gas accretion. It is well known that
bars provoke their own destruction, by driving gas towards 
the center (e.g. Hasan \& Norman 1990). 
After 5 Gyr of galaxy disk evolution, there should not be any bar left.
Why do we observe so many bars today? more than 2/3 of galaxies
are barred. There are even more bars detected 
in NIR images, when dust extinction does not blur them.

The only way to renew a bar in a galaxy is to accrete
external cold gas. The gas will increase the disk mass,
and reduce the bulge-to-disk mass ratio, destabilizing the disk again.
Several bar episodes can process in a Hubble time
(cf Fig 1; Bournaud \& Combes 2002).

We have recently quantified the frequency of bars, in a
sample of 163 galaxies imaged in the NIR by Eskridge et al. (2002).
The gravity torques have been estimated by computing first
the potential of the galaxy, through a 
Fourier transform of its NIR image. 
The intensity $Q_2$ of the $m=2$ torques is defined by 
the maximum over the disk of the ratio 
$F_t(m=2)/F_r$, with $F_t$ the tangential force, and $F_r$ the radial force.
Similarly, the total torques are estimated by the maximum
of $F_t/F_r$, with all $m$ included (Block et al. 2002).
The surprising result is the dearth of axisymmetric galaxies
(cf Fig. 2).  Simulations of external gas accretion
have been made in a sample of objects very similar in types to
the observed sample. The frequency of bars was estimated in
the simulated sample in the same way as for the observed sample.
Comparison of the two histograms of torque intensity allows a
quantification of the accretion rate (Fig. 2). It reveals
that a galaxy doubles its mass in about 10 Gyr, through cold
gas accretion.

\begin{figure}[t]
\centerline{
\psfig{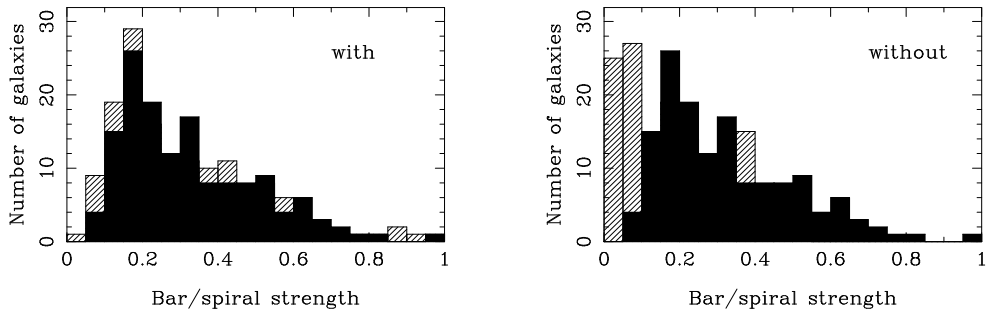}
}
\caption{ Histograms of the bar strength Q$_b$,
defined as the ratio of maximum $m=2$ tangential force to radial force.
{\bf Left} The black histogram is the observed one, and the grey is
from a model with gas accretion, the galaxy doubling its mass in 10 Gyr. 
{\bf Right} The black histogram is the observations, and the grey
is now the model without gas accretion.
(from Block et al. 2002).}
\label{fig2}
\end{figure}

\subsection{Avoidance of dynamical friction}

If the gas is condensed very early in galaxies, and
is tightly bound to massive entities, it will experience
strong dynamical friction at each merging stage,
against the dark matter particles. A way to avoid this
is to eject hot gas in supernovae feedback processes,
but this might be insufficient (e.g. Gnedin \& Zhao 2002,
Maller \& Dekel 2002).

Another scenario is that the gas accretes slowly
in a cold phase onto galaxies,
the hierarchical merging will lose less angular momentum
through dynamical friction (cf. Fig 3).
Gas in this scenario is only slightly bound to galaxies,
and does not form stars too early. Effective accretion 
occurs later. In a merging, the gas is rapidly stripped, 
and does not experience friction.

\begin{figure}[t]
\centerline{
\psfig{figure=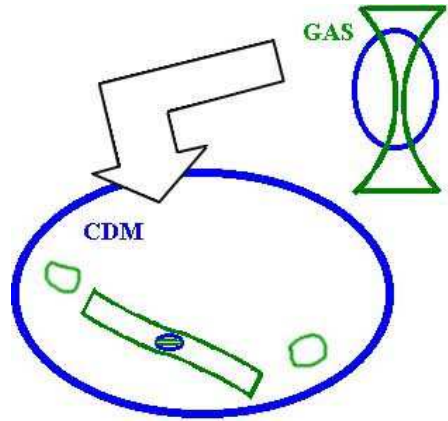,width=7cm,angle=0}
}
\caption{ Schema indicating how the angular momentum loss
can be avoided in mergers of galaxies. A large part of the 
gas mass in the infalling disk at right
is cold and loosely bound in the outer parts
of the galaxy. It is stripped away in the beginning of
the interaction, and as very small clouds, does not experience
the dynamical friction like the baryons bound to the galaxy.}
\label{fig3}
\end{figure}

\section{Disruption of small structures}

CDM models predict for a large-scale halo thousands of
substructure clumps, which at cluster scale correspond
to galaxies and are compatible
with observations. However, this substructure
feature is also expected at galactic scale, and  raises
many problems, when applied for example to our Galaxy.
These small objects are not observed, and could be
only dark halos (Chiu et al 2001), but then would
yield to an excessive heating if present (Moore et al 1999).
The assumption of some warm dark matter (WDM)
helps to reduce the problem (Colin et al 2000).
Sub-structures do form in WDM models, but with
an average concentration parameter which is approximately 
twice as small as that of the corresponding 
CDM sub-halos. This difference in concentration leads to a higher
satellite destruction rate in WDM models, and a better agreement with
observations.
Models of self-interacting dark matter (SIDM), with primordial 
velocity dispersion also reduce substructure and eliminate
central cusps (Hogan \& Dalcanton 2000), but at the expense of 
new problems and fine tuning.

In the scenario where baryons dominate
the mass in the center of galaxies, there is 
more cold gas in dwarf haloes, and
much less mass concentration.
Baryonic clumps heat dark matter through
dynamical friction and smooth any cusp
in dwarf galaxies (El-Zant et al. 2003, in prep).

The material inside galaxies is more dissipative, 
more resonant, and more prone to disruption and merging.
The high merging rate may change the mass function for 
low-mass galaxies (e.g. Mayer et al. 2001).
The disruption of dwarf satellites explain the
formation of the stellar halo of the Milky Way 
(Bullock et al. 2001).
The solution of the cusp problem, leading to more
fragile galaxies, easier to disrupt, can lead also 
to the solution of the substructure problem.

\section{Dark Matter in galaxy clusters}

The fact that the baryonic gas could be highly condensed in galaxies
is supported by the high baryon fraction in clusters.
In rich clusters, the mass of the hot gas in the 
intra-cluster medium (ICM) can be much larger
than the stellar mass in galaxies. Most of the baryons are
in the hot ICM, which represents nearly the baryon fraction 
of the matter in the universe $f_b = \Omega_b/\Omega_m \sim 0.16$.
The relative radial distribution of dark and visible matter is reversed,
the mass becomes more and more visible with radius, contrary
to what occurs in galaxies
(David et al. 1995, Ettori \& Fabian 1999, Sadat \& Blanchard 2001).

Moreover, the hot gas has a relatively high metallicity (1/3 solar), 
and most of the metals in a cluster comes from the
ICM: there is about twice more Fe in the ICM than in galaxies
(Renzini 1997, 2003).

Since the metals are synthetized in galaxies, this means that either
part of the hot gas comes directly from galaxies (by stripping,
or disruption), or that stellar winds and supernovae have enriched the
ICM. In fact, both sources of metals must be there, since metals expelled by
SNe would not be sufficient.
There is the same M$_{Fe}$/L$_B$ in clusters and
galaxies, implying the same processing in
clusters than in the field.
Clusters have not lost iron, nor accreted pristine material
since the ratio between $\alpha$ elements (made in SNII) and iron is
about solar in the ISM of all clusters (with no or little
variation from cluster to cluster). This means that the metals come
from normal stellar nucleosynthesis, and observations are compatible
with the origin of the gas from the outskirts of galaxies.

\section{Conclusion}

The physics of the baryonic gas plays a crucial role in the
formation of galaxies.
The usual assumption that gas is shock heated to the virial temperature
of the dark haloes might not be true.
Instead, there could be cold gas accretion onto galaxies,
with the consequence of more baryons
accreted at a later time, and generally, more baryons condensed
in galaxies, including dark baryons (Valageas et al. 2002). 
The main consequences are:
\begin{itemize}
\item baryons dominate in the center of galaxies, masking the dark matter cusps,
\item existence of large gas extent around galaxies, and less angular momentum lost
by dynamical friction,
\item small galaxies are more prone to disruption and merging, 
which reduces substructures within massive galaxies.
\end{itemize}

\small{

}

\begin{references}
\vspace{-2mm}
\reference Abadi, M. G., Navarro, J. F., Steinmetz, M., Eke, V.R.: 2003, ApJ 591, 499
\reference Afonso, C., Albert J.N., Andersen J. et al.: 2003, A\&A 400, 951
\reference Alcock, C., Allsman, R.A., Alves, D.R. et al.: 2001, ApJ 550, L169
\reference Binney, J.: 2003, MNRAS, in press (astro-ph/0308172)
\reference Binney, J., Gerhard, O., Silk, J.: 2001, MNRAS 321, 471
\reference Block, D., Bournaud, F., Combes, F.. Puerari, I., Buta, R.: 2002, A\&A 394, L35
\reference Bournaud, F., Combes, F.: 2002, A\&A 392, 83
\reference Bullock, J. S., Kravtsov, A. V., Weinberg, D. H.: 2001, ApJ 548, 33
\reference Chiu, W. A., Gnedin, N. Y., Ostriker, J. P.: 2001, ApJ 563, 21
\reference Colín, P., Avila-Reese, V., Valenzuela, O.: 2000, ApJ 542, 622
\reference Combes, F.: 2002 NewAR 46, 755
\reference David, L. P., Jones, C., Forman, W.: 1995, ApJ 445, 578
\reference de Blok, W. J. G., Bosma, A., McGaugh, S.: 2003, MNRAS 340, 657
\reference Eke, V., Efstathiou, G., Wright, L.: 2000, MNRAS 315, L18
\reference El-Zant A., Hoffman Y., Primack J., Combes F., Shlosman I.: 2003, ApJL, 
submitted (astro-ph/0309412)
\reference Eskridge, P. B., Frogel, J. A., Pogge, R. W. et al. 2002, ApJS 143, 73-112
\reference Ettori, S., Fabian, A. C.: 1999 MNRAS 305, 834
\reference Fukushige, T., Makino, J.: 2003, ApJ 588, 674
\reference Gardner, J. P., Katz, N., Hernquist, L., Weinberg, D. H.: 2003, ApJ 587, 1
\reference Katz, N., Keres, D., Dave, R., Weinberg, D.H.: 2002,  in The IGM/Galaxy Connection: 
   Kluwer, p. 185 (astro-ph/0209279)
\reference Knebe, A., Devriendt, J. E. G., Mahmood, A., Silk, J.: 2002, MNRAS 329, 813
\reference Gnedin O.Y., Zhao H.: 2002, MNRAS 333, 299
\reference Hasan, H., Norman, C.: 1990 ApJ 361, 69
\reference Hoekstra, H., van Albada, T. S., Sancisi, R.: 2001, MNRAS 323, 453
\reference Hogan, C. J., Dalcanton, J. J.: 2000, Phys Rev D 62, 3511
\reference Lassere, T., Afonso, C., Albert J.N. et al. : 2000, A\&A 355, L39
\reference Maller A.H., Dekel A.: 2002, MNRAS 335, 487
\reference Mayer, L., Governato, F., Colpi, M. et al. 2001 ApJ 547, L123
\reference McGaugh, S. S., Barker, M. K., de Blok, W. J. G.: 2003, ApJ 584, 566
\reference McGaugh, S. S., Schombert, J. M., Bothun, G. D., de Blok, W. J. G.:
   2000, ApJ 533, L99    
\reference Meza, A., Navarro, J. F., Steinmetz, M., Eke, V. R.: 2003, ApJ 590, 619
\reference Milosavljevic, M., Merritt, D.: 2001, ApJ 563, 34
\reference Mo, H. J., Mao, S., White, S. D. M.: 1998 MNRAS 295, 319
\reference Moore, B., Ghigna, S., Governato, F., et al.: 1999, ApJ 524, L19
\reference Navarro, J.F., Frenk, C.S., White, S.D.M.: 1997, ApJ 490, 493
\reference Navarro, J. F., Steinmetz, M.: 2000, ApJ 538, 477
\reference Pfenniger, D., Combes, F., Martinet, L.: 1994, A\&A 285, 79
\reference Pfenniger, D., Combes, F.: 1994, A\&A 285, 94
\reference Renzini, A.: 1997, ApJ 488, 35
\reference Renzini, A.: 2003,
   in Clusters of Galaxies: Probes of Cosmological Structure and Galaxy Evolution,
   Carnegie Observatories Symposium III. (astro-ph/0307146)
\reference Sadat, R., Blanchard, A.: 2001, A\&A 371, 19
\reference Sand, D.J., Treu, T., Ellis, R.S.: 2002, ApJ 574, L129
\reference Spergel, D. N., Steinhardt, P. J.: 2000, Phys Rev Let 84, 3760
\reference Swaters, R. A., Madore, B. F., van den Bosch, F. C., Balcells, M.: 
   2003, ApJ 583, 732
\reference Thacker, R. J., Couchman, H. M. P.: 2001, ApJ 555, L17
\reference Valageas, P., Schaeffer, R., Silk, J.: 2002, A\&A 388, 741
\reference Weil M.L., Eke, V. R., Efstathiou, G.: 1998, MNRAS 300, 773
\reference Weinberg, M. D., Katz, N.: 2002, ApJ 580, 627
\end{references}
\end{document}